\documentclass[12pt]{article}

\usepackage[T1]{fontenc}
\usepackage[utf8]{inputenc}
\usepackage{textcomp}

\PassOptionsToPackage{unicode}{hyperref}
\PassOptionsToPackage{hyphens}{url}
\PassOptionsToPackage{dvipsnames,svgnames,x11names}{xcolor}

\usepackage{amsmath, amssymb, amsfonts, mathrsfs}
\usepackage{bm}         % 数学粗体
\usepackage{bbm}        % 黑板粗体
\usepackage{dsfont}     % 双线字体

\usepackage{graphicx}
\usepackage{xcolor}
\usepackage{booktabs}
\usepackage{multirow}
\usepackage{array}
\usepackage{enumerate}
\usepackage{comment}
\usepackage{setspace}
\usepackage{calc}
\usepackage{etoolbox}

\usepackage{natbib}

\usepackage{caption}
\usepackage{subcaption}
\usepackage{longtable}
\usepackage{footnote}
\makesavenoteenv{longtable}

\usepackage{lmodern}
\usepackage{microtype}
\UseMicrotypeSet[protrusion]{basicmath}

\newtheorem{theorem}{Theorem}[section]

\newtheorem{proposition}{Proposition}[section]

\newtheorem{definition}{Definition}[section]

\makeatletter
\def\maxwidth{\ifdim\Gin@nat@width>\linewidth\linewidth\else\Gin@nat@width\fi}
\def\maxheight{\ifdim\Gin@nat@height>\textheight\textheight\else\Gin@nat@height\fi}
\makeatother
\setkeys{Gin}{width=\maxwidth,height=\maxheight,keepaspectratio}

\makeatletter
\def\fps@figure{htbp}
\def\fps@table{htbp}
\makeatother

\usepackage{hyperref}
\usepackage{xurl}
\usepackage{bookmark}

\hypersetup{
  pdftitle={Title},
  pdfauthor={Author 1; Author 2},
  pdfkeywords={3 to 6 keywords, that do not appear in the title},
  colorlinks=true,
  linkcolor=blue,
  filecolor=Maroon,
  citecolor=blue,
  urlcolor=blue,
  pdfcreator={LaTeX}
}

\newcommand{\anon}{1}

\begin{document}

\def\spacingset#1{\renewcommand{\baselinestretch}%
{#1}\small\normalsize} \spacingset{1}

%%%%%%%%%%%%%%%%%%%%%%%%%%%%%%%%%%%%%%%%%%%%%%%%%%%%%%%%%%%%%%%%%%%%%%%%%%%%%%

\if1\anon
{
  \title{\bf Fused Spatial Latent Block Models for Co-Clustering}
  \author{Biao Cai 1%\thanks{The authors gratefully acknowledge \textit{please remember to list all relevant funding sources in the version that gives all author information}} 
  \thanks{The authors are listed in the alphabetic order}\hspace{.2cm}\\
    Department of Decision Analytics and Operations, City University of Hong Kong\\
    and \\
    Yuanxing Chen 2\thanks{Corresponding author. Email: yxchen\_research@163.com.}\\
    School of Economics \& Management, Fuzhou University\\
    and \\
    Kuangnan Fang 3 \\
    School of Economics, Xiamen University\\
    and \\
    Xiaolong Lin 4 \\
    School of Economics, Xiamen University}
  \maketitle
} \fi

\if0\anon
{
  \bigskip
  \bigskip
  \bigskip
  \begin{center}
    {\LARGE\bf Fused Spatial Latent Block Models for Co-Clustering}
\end{center}
  \medskip
} 
\fi

\bigskip
\begin{abstract}

Spatial transcriptomics is a rapidly growing technique that captures gene expression together with spatial coordinates in intact tissue sections, enabling in situ mapping of transcriptional activity.
This technology offers unprecedented opportunities to study tissue heterogeneity and spatial gene expression patterns. 
Uncovering the associations between spatially variable gene modules and spot types can advance our understanding of pathological mechanisms. 
However, rigorous statistical methods that exploit spatial information to achieve spatially coherent co-clustering of spots and genes are still lacking, and theoretical investigations in this direction remain limited.

We propose a fused spatial latent block model (F-SpLBM).
Our model uses the LBM to uncover co-expression patterns between spots and genes, penalized fusion to automatically determine the number of co-clusters, and the Potts model to incorporate spatial information.
We establish that the fusion-based procedure recovers the true block structure with the misclassification rate converging at a super-polynomial rate. 
We also prove asymptotic normality of the parameter estimators and quantify the accuracy gain from spatial smoothing.
Simulations and real-data analyses demonstrate that F-SpLBM yields spatially coherent and biologically interpretable clustering results.

\end{abstract}

\noindent%
\vfill

\newpage
\spacingset{1.8} % DON'T change the spacing!

\section{Introduction}

Spatial transcriptomics (ST) is a recently developed biological technology. 
It captures data from an array of spots on intact tissue sections, and measures gene expression together with its spatial location simultaneously \citep{stahl2016spatial}. 
Popular RNA profiling techniques such as RNA-seq and single-cell RNA-seq lack this spatial information. 
Therefore, ST provides a new way to investigate the spatial organization and function of complex tissues, including identifying rare cell populations, reconstructing differentiation trajectories, and characterizing the spatial heterogeneity inherent in these tissues \citep{trapnell2014dynamics, kim2015single, notta2016distinct, marx2021method}.

\begin{figure}[htbp]
    \centering
    \includegraphics[width=0.6\textwidth]{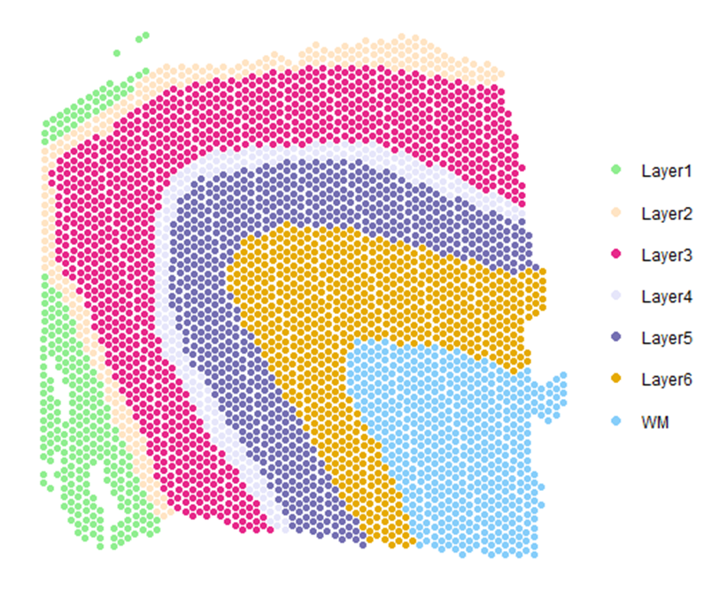} 
    \caption{A tissue sample from the human dorsolateral prefrontal cortex. Dots represent individual spots, and different colors denote the cluster labels manually annotated by \cite{maynard2021transcriptome}.}
    \label{fig:spatial_smoothness}
\end{figure}

An important task of ST data analysis is to reveal tissue heterogeneity, that is, to partition spots into disjoint clusters that reflect different histological functions. 
However, this task faces two constraints. 
Morphologically, spots within the same cluster tend to form contiguous tissue domains and show spatial smoothness, as illustrated in Figure \ref{fig:spatial_smoothness}. 
Ignoring spatial smoothness and relying only on gene expression risks creating fragmented, unnatural clusters. 
From the bioscience perspective, identifying associations between gene modules and spot types can help us better understand pathological mechanisms \citep{higgins2025stihc}. 
Therefore, we consider a spatial co-clustering problem that considers tissue heterogeneity, spatial smoothness, and co-patterns of genes and spots.

%High-dimensional data clustering has been extensively studied, with three prevailing approaches being sparse clustering \citep{witten2010framework, yao2025bayesian}, dimensionality reduction-based clustering \citep{yang2022sc} and co-clustering \citep{biernacki2023survey}. 
%Sparse clustering extends conventional methods by imposing penalties on features, thereby achieving simultaneous variable selection and clustering. 
%While this approach is well-suited for data with numerous noisy features, it is less appropriate for ST data, where most genes carry meaningful biological signal rather than constituting pure noise. 
%The second approach first reduces dimensionality via techniques such as principal component analysis and then performs clustering in the low-dimensional space. 
%However, it suffers from the limitations: the low-dimensional embedding may not meaningfully correspond to the underlying clustering structure, and a systematic theoretical framework to guarantee this correspondence remains absent. 
%Unlike these two approaches, co-clustering retains the original data matrix. By simultaneously partitioning both rows and columns, it uncovers homogeneous submatrix blocks. 
%This capacity to reveal associative patterns between gene modules and spot types renders co-clustering especially attractive for the analysis of ST data.

The Latent Block Model (LBM, \cite{govaert2003clustering}) provides a solid probabilistic framework for co-clustering. 
Existing studies have investigated the theoretical properties of the LBM. 
\cite{brault2020consistency} proves consistency and asymptotic normality for parameter estimation in the univariate case. 
\cite{zhao2024variational} established vanishing misclassification rates for clustering performance under the Poisson distribution assumption. 
However, ST data from popular platforms such as 10X Genomics Visium are obtained by standardizing and log-transforming the unique molecular identifier (UMI) counts, and are often modeled under a normal distribution \citep{love2014moderated, hafemeister2019normalization}. 
This gap between Poisson-based theory and normal-based practice calls for an extension of existing theoretical results to the normal and multivariate setting. 

Several statistical methods have been proposed to incorporate spatial information when clustering ST data. 
SpaRTaCo \citep{sottosanti2023co} introduces a spatial covariance structure. 
However, its estimation relies on the Metropolis-Hastings algorithm, and it cannot guarantee spatial smoothness in the clustering results. 
A more direct approach is to encode spatial dependence through Markov random fields (MRFs), which regularize the cluster labels of neighboring spots. 
MRF-based approaches have shown strong empirical performance in clustering ST data \citep{liu2022joint, yan2024bayesian}. 
The Potts model \citep{wu1982potts} is especially attractive: it is computationally convenient and promotes cluster smoothness by penalizing label discontinuities across adjacent spots. 
In this article, we integrate the Potts model into the LBM framework, providing both theoretical guarantees and spatial smoothness for tissue region segmentation. 

A popular approach to determining the number of co-clusters is to perform an exhaustive search over row and column cluster counts, maximizing the integrated completed log-likelihood (ICL; \cite{biernacki2000assessing}). 
However, this searching is especially time-consuming for large datasets. 
Motivated by \cite{manole2021estimating}, we present a penalized fusion to automatically determine co-clusters, and show that the fusion step converges in a single iteration for adequate sample sizes, yielding substantial computational savings.
Because evaluating the joint posterior distribution of row and column cluster labels is NP-hard, the usual E-step is not intractable for the LBM \citep{biernacki2023survey}. 
We develop a variational expectation-maximization (VEM) algorithm to estimate the parameter. 
Our procedure first estimates the LBM with an overspecified number of co-clusters using VEM algorithm, then fuses row and column clusters with similar centers through an adaptive penalty. 
Finally, we apply a spatial smoothing adjustment via the Potts model.

We makes several theoretical contributions spanning parameter estimation, clustering accuracy, and the recovery of the co-cluster structure. 
To the best of our knowledge, we are the first to establish theoretical properties of the VEM algorithm under overspecification. 
Specifically, we prove that the generalized misclassification rate of \cite{zhang2021label} converges to zero at a super-polynomial rate. 
Unlike the case of one-way clustering, we demonstrate that the probability of producing spurious small clusters with near-zero proportion converges asymptotically to zero. 
We also establish the consistency of the parameter estimates. 
In the fusion step, the adaptive weights are chosen to guarantee exact block recovery and the asymptotic normality of the parameter estimators. 
Finally, we explicitly quantify the improvement in clustering accuracy gained from incorporating spatial smoothness. 
Taken together, these results provide a rigorous theoretical foundation for spatial co-clustering and substantially extend the applicability of the LBM framework.

The rest of this paper is organized as follows. Section \ref{Methodology} introduces the spatial LBM and the proposed VEM algorithm together with penalized fusion. 
Section \ref{Estimation} describes the computational implementation. 
Section \ref{Theoretical_Properties} establishes statistical properties of the estimator. Section \ref{Simulation} reports extensive simulation studies under various scenarios. 
Section \ref{Real_Data} demonstrates the practical utility of the method through a real-data application.

\section{Methodology}\label{Methodology}

\begin{figure}[htbp]
    \centering
    \includegraphics[width=0.35\textwidth]{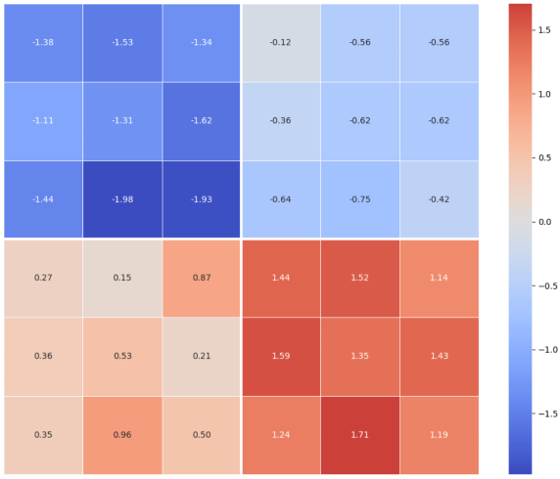} 
    \caption{A heatmap of data with latent block structure.}
    \label{fig:methodology_structure}
\end{figure}

Let $\mathbf X = (x_{ij}) \in \mathbb{R}^{n \times d}$ denote the gene expression matrix, where each row corresponds to a spatial spot and each column to one of the $d$ measured genes.
We assume that $\mathbf X$ follows a latent block structure with $K$ row clusters and $L$ column clusters, resulting in $K \times L$ co-clusters (Figure \ref{fig:methodology_structure} provides a schematic illustration). 
Let $z_i \in [K]$ and $w_j \in [L]$ denote the row- and column-cluster labels, and define the one-hot indicators $z_{ik} = \mathbbm{1}(z_i = k)$ and $w_{jl} = \mathbbm{1}(w_j = l)$.
Conditional on $(\bm z, \bm w)$, the entries $x_{ij}$ are assumed to be mutually independent with Gaussian densities $x_{ij} \mid (z_{ik}=1, w_{jl}=1) \sim \mathcal{N}(\mu_{kl}, \lambda_{kl})$, where $\mu_{kl}$ and $\lambda_{kl}$ are the block-specific mean and variance.
We collect the mean and variance parameters into matrices $\bm \mu = (\mu_{kl}) \in \mathbb{R}^{K \times L}$ and $\bm \Lambda = (\lambda_{kl}) \in \mathbb{R}^{K \times L}$.
The row and column labels are a priori independent with multinomial distributions
\[
p_z(\bm z) = \prod_{i=1}^n \prod_{k=1}^K \pi_k^{z_{ik}}, \quad
p_w(\bm w) = \prod_{j=1}^d \prod_{l=1}^L \rho_l^{w_{jl}},
\]
where $\bm\pi = (\pi_1,\dots,\pi_K)^\top$ and $\bm\rho = (\rho_1,\dots,\rho_L)^\top$ are probability vectors.

\subsection{Variational Likelihood}

Since the cluster labels are unobserved, maximum likelihood estimation falls into the framework of missing-data problems. 
The EM algorithm provides a natural framework in such settings. 
By Jensen's inequality, we have
\begin{equation*}
    \ln f(\mathbf{X}) = \ln \sum_{\bm z\in[K]^n}\sum_{\bm w\in[L]^d} f(\mathbf{X}, \bm z,\bm w)
    \ge \sum_{\bm z\in[K]^n}\sum_{\bm w\in[L]^d} q(\bm z,\bm w)\ln \frac{f(\mathbf{X}, \bm z,\bm w)}{q(\bm z,\bm w)},
\end{equation*}
where $q$ is any probability measure over $[K]^n\times[L]^d$, and equality holds if $q(\bm z,\bm w) = \mathbb{P}(\bm z,\bm w\mid\mathbf{X})$. 
Evaluating the exact posterior in the E-step and calculating $\sum_{\bm z\in[K]^n}\sum_{\bm w\in[L]^d}$ are intractable. 
A standard remedy is variational inference. 
By imposing a mean-field factorizability constraint $q(\bm z,\bm w) = q_z(\bm z)q_w(\bm w)$, we obtain the evidence lower bound (ELBO):
\begin{equation*}
    F(\bm\Theta) = \max_{q_z,q_w}\sum_{\bm z\in[K]^n}\sum_{\bm w\in[L]^d} q_z(\bm z)q_w(\bm w)\ln \frac{f(\mathbf{X}, \bm z,\bm w)}{q_z(\bm z)q_w(\bm w)},
\end{equation*}
where $\bm\Theta = (\bm\pi^\top,\bm\rho^\top, \mathrm{vec}(\bm \mu)^\top, \mathrm{vec}(\bm\Lambda)^\top)^\top$. 
As we will establish in Section \ref{Theoretical_Properties}, the cluster assignments are asymptotically consistent, implying that the true posterior $\mathbb{P}(\bm z,\bm w\mid\mathbf{X})$ concentrates on a single configuration. 
This asymptotic concentration provides a theoretical justification for the mean-field approximation: when the posterior is sharply peaked, the factorized surrogate incurs negligible approximation error.

The algorithm alternates between obtaining the ELBO (E-step) and maximizing $F(\bm\Theta)$ with respect to $\bm\Theta$ (M-step). 
In the E-step, the variational distributions take the form $q_z(\bm z)=\prod_{i=1}^n\prod_{k=1}^K s_{ik}^{I(z_i=k)}$ and $q_w(\bm w)=\prod_{j=1}^d\prod_{l=1}^L t_{jl}^{I(w_j=l)}$, where $\{s_{ik}\}$ and $\{t_{jl}\}$ are the variational probabilities. 
Substituting these into the ELBO leads to the maximization of
\begin{equation}
    \begin{aligned}
    \mathcal J(\bm s, \bm t, \bm\Theta) =& -\sum_{i=1}^{n}\sum_{k=1}^{K} s_{ik} \ln\frac{s_{ik}}{\pi_{k}} - \sum_{j=1}^{d}\sum_{l=1}^{L} t_{jl} \ln\frac{t_{jl}}{\rho_l}\\
   & +\sum_{i=1}^{n}\sum_{k=1}^{K}\sum_{j=1}^{d}\sum_{l=1}^{L} s_{ik}t_{jl} \ln f(x_{ij};\mu_{kl},\lambda_{kl}).
    \end{aligned}
    \label{J}
\end{equation}

\subsection{Penalized Fusion}

Determining the true number of co-clusters is a key problem in model-based clustering. 
The penalized fusion framework merges clusters whose centers coincide because they represent the same components \citep{manole2021estimating}. 
This principle leads us to impose the following penalty to merge clusters with similar centers: 
\begin{equation}\label{penalty}
    \mathcal{P}(\bm\mu) = \gamma\sum_{k_1<k_2} w^{(r)}_{k_1k_2}\big\| \bm\mu_{k_1\cdot} - \bm\mu_{k_2\cdot}\big\| + \gamma\sum_{l_1<l_2} w^{(c)}_{l_1l_2}\big\| \bm\mu_{\cdot l_1} - \bm\mu_{\cdot l_2}\big\|,
\end{equation}
where the weights ${ w^{(r)}_{k_1k_2}}$ and ${ w^{(c)}_{l_1l_2}}$ are chosen adaptively, and $\gamma$ is a tuning parameter that controls the strength of fusion. 
Its value is selected by maximizing the ICL of the form
\begin{equation*}
    \mathrm{ICL} = \ln f(\mathbf X,\widehat{\bm z}, \widehat{\bm w}; \widehat{\bm\Theta}) - \frac{K_\gamma-1}{2}\ln n - \frac{L_\gamma-1}{2}\ln d-K_\gamma L_\gamma\ln(nd).
\end{equation*}

\subsection{Spatial Lattice System}
\begin{figure}[htbp]
    \centering
    \includegraphics[width=0.35\textwidth]{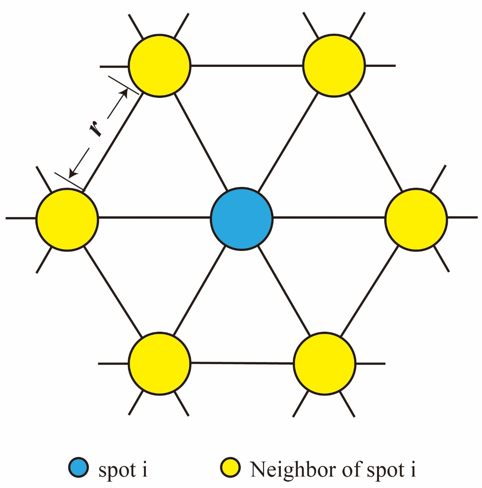} 
    \caption{Lattice patterns and adjacency relationships for organized spots.}
    \label{fig:methodology_neighbor}
\end{figure}

We now consider a lattice system, a standard setup in spatial transcriptomics research. 
Suppose the $n$ spatial spots lie on a tissue section and are arranged on a regular triangular lattice with side length $r$. 
Each spot is associated with a spatial coordinate $\bm \xi_i = (\xi_{i1}, \xi_{i2})^\top$. 
The neighborhood of spot $i$ is defined as $\mathcal N_i = \{j : \|\bm \xi_i - \bm \xi_j\| = r\}$, that is, the set of spots located at Euclidean distance $r$ from spot $i$ (Figure \ref{fig:methodology_neighbor}). 

We use a Potts model to capture spatial dependence among neighboring spots:
\begin{equation}\label{Potts}
 p_z(\bm z) = C_{\beta}^{-1} \exp\left\{ \frac{1}{2}\beta\sum_{i=1}^{n} \sum_{j\in\mathcal N_i}\mathbbm 1(z_i = z_j) \right\},
\end{equation}
where $\beta \ge 0$ and $C_{\beta}$ is the intractable normalizing constant. A larger $\beta$ imposes a stronger penalty on label mismatches between neighbors, thus promoting greater spatial smoothness.

We estimate the interaction parameter $\beta$ by first obtaining preliminary row-cluster assignments from a non-spatial LBM, and then maximizing the pseudo-likelihood of the Potts model. 
With this estimated $\beta$, we re-estimate the full model, updating the variational probabilities of the row labels. 
As we show in the subsequent theoretical analysis, simulations, and real-data experiments, this spatial refinement step improves both clustering accuracy and spatial smoothness, with particularly pronounced gains in small-sample and low signal-to-noise settings.

\section{Estimation}\label{Estimation}
Our procedure first estimates an LBM with an overspecified number of co-clusters using the VEM algorithm and obtains the weights for the subsequent penalties. 
Then it fuses row and column clusters with similar centers through an adaptive penalty. 
Finally, we apply a spatial smoothing adjustment via the Potts model.

\textbf{Step I. VEM algorithm.}
We solve \eqref{J} with the numbers of row and column clusters, $K$ and $L$, intentionally chosen to exceed the true values. 

In the E-step, given $\bm\Theta^{(m)}$, and $\bm t^{(m)}$, we update $\bm s$ via $s_{ik}^{(m+1)} \propto \pi_{k}^{(m)} A_{ik}^{(m)}$ (normalized over $k$), and then update $\bm t$ with $t_{jl}^{(m+1)} \propto \rho_l^{(m)} B_{jl}^{(m)}$ (normalized over $l$), where $\ln A_{ik}^{(m)} = - \frac{1}{2}\sum_{j,l} t_{jl}^{(m)} \ln f(x_{ij};\mu_{kl}^{(m)},\lambda_{kl}^{(m)})$, and $\ln B_{jl}^{(m)} = - \frac{1}{2}\sum_{i,k} s_{ik}^{(m+1)} \ln f(x_{ij};\mu_{kl}^{(m)},\lambda_{kl}^{(m)})$.  

In the M-step, for all $k\in[K]$ and $l\in[L]$, the update for $\pi_k$ and $\rho_l$ are given by $\pi_k^{(m+1)} = \frac{1}{n}\sum_{i=1}^n s_{ik}^{(m+1)}$ and $\rho_l^{(m+1)} = \frac{1}{d}\sum_{j=1}^{d}t_{jl}^{(m+1)}$, and for $(\mu_{kl},\lambda_{kl})$ are:
\begin{align}
    \mu_{kl}^{(m+1)} =& \frac{\sum_{i=1}^{n}\sum_{j=1}^{d} s_{ik}^{(m+1)} t_{jl}^{(m+1)} x_{ij}}{\sum_{i=1}^{n}\sum_{j=1}^{d} s_{ik}^{(m+1)} t_{jl}^{(m+1)} }, \\
    \lambda_{kl}^{(m+1)} =& \frac{\sum_{i=1}^{n}\sum_{j=1}^{d} s_{ik}^{(m+1)} t_{jl}^{(m+1)} (x_{ij}-\mu_{kl}^{(m+1)})^2}{\sum_{i=1}^{n}\sum_{j=1}^{d} s_{ik}^{(m+1)} t_{jl}^{(m+1)} }.
\end{align}

Let $\mathcal{J}^{(m)} = \mathcal J(\bm s^{(m)}, \bm t^{(m)}, \bm\Theta^{(m)})$. The VEM algorithm iterates until $|(\mathcal{J}^{(m+1)} - \mathcal{J}^{(m)})/\mathcal{J}^{(m)}|\le\varepsilon$, where $\varepsilon$ is the relative tolerance. Let $({\bm s}^{(M)}, {\bm t}^{(M)}, {\bm\Theta}^{(M)})$ be the results of VEM algorithm.
\begin{proposition}
    The sequence $\{({\bm s}^{(m)}, {\bm t}^{(m)}, {\bm\Theta}^{(m)})\}_{m\ge0}$ generated by VEM algorithm converges to a stationary point of $\mathcal{J}(\bm s, \bm t, \bm\Theta)$.
\end{proposition}
This result ensures that the algorithm converges from any valid initialization, ruling out divergence in practical implementation.

\textbf{Step II. penalized fusion.} 
 Following \cite{zou2006adaptive}, we construct adaptive weights that inflate the penalty for closely spaced cluster centers: 
\begin{equation}\label{weights}
    w^{(r)}_{k_1k_2} = \frac{\sqrt{L}}{\| {\bm{\mu}}^{(M)}_{k_1\cdot} - {\bm{\mu}}^{(M)}_{k_2\cdot} \|},\quad w^{(c)}_{l_1l_2} = \frac{\sqrt{K}}{\| {\bm{\mu}}^{(M)}_{\cdot l_1} - {\bm{\mu}}^{(M)}_{\cdot l_2} \|},
\end{equation}
Fixing the variational posteriors at $s^{(M)}$ and $t^{(M)}$, we minimize the following penalized objective with respect to $(\mu, \Lambda)$:
\begin{equation}\label{penalized_objective_function}
    \mathcal Q(\bm \mu, \bm\Lambda) =\frac{1}{2nd}\sum_{i=1}^{n}\sum_{k=1}^{K}\sum_{j=1}^{d}\sum_{l=1}^{L} s^{(M)}_{ik} t^{(M)}_{jl} \left[ \ln\lambda_{kl} + \frac{(x_{ij}-\mu_{kl})^2}{\lambda_{kl}} \right] + \mathcal{P}(\bm\mu). 
\end{equation}
Let $\{\widetilde{\bm\mu},\widetilde{\bm\Lambda}\} = \arg\min_{\bm\mu,\bm\Lambda} \mathcal Q(\bm \mu, \bm\Lambda)$. To optimize \eqref{penalized_objective_function}, we adopt the alternating direction method of multipliers (ADMM).
In particular, let $\bm h_{k_1k_2} = \bm\mu_{k_1\cdot} - \bm\mu_{k_2\cdot}$ and $\bm v_{l_1l_2} = \bm\mu_{\cdot l_1} - \bm\mu_{\cdot l_2}$.
Let $\bm a = \{ \bm a_{k_1k_2}\}$ and $\bm b = \{\bm b_{l_1l_2}\}$ be the Lagrange multipliers, and $\varphi>0$ be the penalty parameter (we set $\varphi = 1$ in implementation). The augmented Lagrangian function is
\begin{align*}
    & \mathcal L(\bm\mu, \bm\Lambda, \bm h, \bm v, \bm a, \bm b) =  \frac{1}{2nd}\sum_{i=1}^{n}\sum_{k=1}^{K}\sum_{j=1}^{d}\sum_{l=1}^{L} s^{(M)}_{ik} t^{(M)}_{jl} \left[ \ln\lambda_{kl} + \frac{(x_{ij}-\mu_{kl})^2}{\lambda_{kl}} \right]\\
    & + \sum_{k_1<k_2}\left[ \gamma w^{(r)}_{k_1k_2}\big\| \bm h_{k_1k_2}\big\| + \bm a_{k_1k_2}^\top(\bm\mu_{k_1\cdot} - \bm\mu_{k_2\cdot} - \bm h_{k_1k_2}) + \frac{\varphi}{2}\big\| \bm\mu_{k_1\cdot} - \bm\mu_{k_2\cdot} - \bm h_{k_1k_2} \big\|^2 \right]\\
    & + \sum_{l_1<l_2}\left[ \gamma w^{(c)}_{l_1l_2}\big\| \bm v_{l_1l_2}\big\| + \bm b_{l_1l_2}^\top(\bm\mu_{\cdot l_1} - \bm\mu_{\cdot l_2} - \bm v_{l_1l_2}) + \frac{\varphi}{2}\big\| \bm\mu_{\cdot l_1} - \bm\mu_{\cdot l_2} - \bm v_{l_1l_2} \big\|^2 \right].
\end{align*}
Then $\mu_{kl}^{(m'+1)} = C_{1,kl}^{(m')}/C_{2,kl}^{(m')}$, where
\begin{align*}
    C_{1,kl}^{(m')} =& \frac{1}{nd\lambda_{kl}^{(m)}}\sum_{i=1}^{n}\sum_{j=1}^{d} s^{(M)}_{ik} t^{(M)}_{jl} x_{ij} + \varphi\sum_{k_1\neq k}\mu_{k_1l}^{(m')} + \varphi\sum_{l_1\neq l}\mu_{kl_1}^{(m')}\\
    & + \sum_{k_1\neq k}\big( \varphi h_{kk_1, l}^{(m')} - a_{kk_1, l}^{(m')} \big) + \sum_{l_1\neq l}\big( \varphi v_{ll_1, k}^{(m')} - b_{ll_1, k}^{(m')} \big),\nonumber\\
    C_{2,kl}^{(m')} =& \frac{1}{nd\lambda_{kl}^{(m')}}\sum_{i=1}^{n}\sum_{j=1}^{d} s^{(M)}_{ik} t^{(M)}_{jl} x_{ij} + \varphi(K + L - 2).
\end{align*}
The update for $\bm\Lambda^{(m'+1)}$ remains the same as before.
The auxiliary variables $\bm h^{(m'+1)}$ and $\bm v^{(m'+1)}$ are updated using the soft-thresholding operator,
\begin{align*}
    \bm h_{k_1k_2}^{(m'+1)} =& \mathrm S_{\gamma w^{(r)}_{k_1k_2}}(\|\bm a_{k_1k_2}^{(m')} + \varphi(\bm\mu_{k_1\cdot}^{(m'+1)} - \bm\mu_{k_2\cdot}^{(m'+1)})\|)\frac{\bm a_{k_1k_2}^{(m')} + \varphi(\bm\mu_{k_1\cdot}^{(m'+1)} - \bm\mu_{k_2\cdot}^{(m'+1)})}{\varphi\|\bm a_{k_1k_2}^{(m')} + \varphi(\bm\mu_{k_1\cdot}^{(m'+1)} - \bm\mu_{k_2\cdot}^{(m'+1)})\|}\\
    \bm v_{l_1l_2}^{(m'+1)} =& \mathrm S_{\gamma w^{(c)}_{l_1l_2}}(\|\bm b_{l_1l_2}^{(m')} + \varphi(\bm\mu^{(m'+1)}_{\cdot l_1} - \bm\mu^{(m'+1)}_{\cdot l_2})\|)\frac{\bm b_{l_1l_2}^{(m')} + \varphi(\bm\mu^{(m'+1)}_{\cdot l_1} - \bm\mu^{(m'+1)}_{\cdot l_2})}{\varphi\|\bm b_{l_1l_2}^{(m')} + \varphi(\bm\mu^{(m'+1)}_{\cdot l_1} - \bm\mu^{(m'+1)}_{\cdot l_2})\|}
\end{align*}
where $S_{\gamma}(x) = \mathrm{sign}(x)(|x|-\gamma)_+$. Finally, the Lagrange multipliers are updated by
\begin{align*}
    \bm a_{k_1k_2}^{(m'+1)} =& \bm a_{k_1k_2}^{(m')} + \varphi(\bm\mu^{(m'+1)}_{k_1\cdot} - \bm\mu^{(m'+1)}_{k_2\cdot} - \bm h_{k_1k_2}^{(m'+1)}) \\
    \bm b_{l_1l_2}^{(m'+1)} =& \bm b_{l_1l_2}^{(m')} + \varphi(\bm\mu^{(m'+1)}_{\cdot l_1} - \bm\mu^{(m'+1)}_{\cdot l_2} - \bm v_{l_1l_2}^{(m'+1)}).
\end{align*}
When $\gamma$ is suitably large, the ADMM yields $\widetilde{h}_{k_1k_2} = \bm0$, signaling that the corresponding row-cluster centers have been fused. Let $(\widetilde{K}, \widetilde{L})$ denote the numbers of clusters after fusion. We then recompute the variational posteriors with the fused parameters to obtain $\widetilde{s}$ and $\widetilde{t}$, and define the hard cluster assignments $\widetilde{z}_i = \arg\max_k \widetilde{s}_{ik}$.

Similarly, we can define $\mathcal{L}^{(m')}$. The penalized fusion algorithm iterates until $|(\mathcal{L}^{(m'+1)} - \mathcal{L}^{(m')})/\mathcal{L}^{(m')}|\le\varepsilon$. 

\textbf{Step III. spatial adjustment.} 
Let $S(z) = \sum_{i=1}^n \sum_{j \in \mathcal{N}_i} \mathbbm{1}(z_i = z_j)$. The maximum likelihood estimator of $\beta$ from \eqref{Potts} satisfies the moment condition $S(\widetilde{z}) = \mathbb{E}_\beta[S(z)]$, where $\widetilde{z}$ are the hard cluster assignments from Step II. Computing the expectation requires summing over $K^n$ configurations, which is infeasible. We therefore resort to the maximum pseudo-likelihood estimator (MPLE), whose consistency for the Potts model is established in \cite{winkler2012image}. The conditional pseudo-likelihood is
\begin{equation}\label{zi}
  \mathbb P(z_i = k\mid \bm z_{-i}) \propto \exp\left\{ \beta\sum_{j\in \mathcal N_i}\mathbbm 1(z_j = k) \right\}.
\end{equation}
Although the MPLE has no closed form, the log-pseudo-likelihood is strictly concave in $\beta$. We maximize it via a one-dimensional grid search:
\begin{equation*}
    \widetilde\beta = \arg\max_{\beta\in \{\beta_1, \beta_2, \cdots, \beta_S\}}\sum_{i=1}^{n}\sum_{k=1}^{K} \widetilde{s}_{ik}\ln\frac{\exp\{\beta \sum_{j\in\mathcal N_i}\mathbbm 1(\widetilde{z}_i = k)\}}{\sum_{k'=1}^{K}\exp\{\beta \sum_{j\in\mathcal N_i}\mathbbm 1(\widetilde{z}_i = k')\}}.
\end{equation*}
We refine the row-cluster posteriors by incorporating the spatial prior: $s_{ik} \propto \pi_{ik} A_{ik}$, where $\pi_{ik} = \mathbb P(z_i=k\mid\widetilde{\bm z}_{\mathcal N_i};\widetilde\beta)$, and $A_{ik}$ is defined in the VEM step.

\section{Theoretical Properties}\label{Theoretical_Properties}
We assume that the numbers of row and column clusters used in the model, $K$ and $L$, are known upper bounds on the true numbers $K_0$ and $L_0$ (that is, $K\ge K_0$, $L\ge L_0$).  
For any model parameter $\alpha$, let $\alpha^*$ denote its true value under correct specification.  
We impose the following regularity conditions.

\begin{enumerate}[(C1)]
  \item There exist positive constants $\pi_{\min}$, $\pi_{\max}$, $\rho_{\min}$, $\rho_{\max}$, $\mu_{\max}$, $\lambda_{\min}$, $\lambda_{\max}$ such that for all $k\in[K_0]$ and $l\in[L_0]$, $\pi_{\mathrm {min}} \le \pi_{k}^*\le \pi_{\mathrm {max}}$, $\rho_{\mathrm {min}} \le \rho_{l}^*\le \rho_{\mathrm {max}}$, $|\mu^*_{kl}|\le \mu_{\mathrm {max}}$, and $\lambda_{\mathrm {min}} \le \lambda^*_{kl}\le \lambda_{\mathrm {max}}$.
  \item The true mean matrix $\bm\mu^*\in\mathbb{R}^{K_0\times L_0}$ has $K_0$ distinct rows and $L_0$ distinct columns.
  \item The dimensions grow at rates satisfying $\ln d = o(n)$ and $\ln n = o(d)$.
\end{enumerate}
Conditions (C1) and (C2) are adapted from \cite{zhao2024variational}, and Condition (C3) follows \cite{brault2020consistency}.  
Condition (C1) ensures that the true parameters lie in a compact set away from the boundaries.  
Condition (C2) guarantees identifiability of $\bm\mu^*$ up to simultaneous permutations of the row and column labels.  
Condition (C3) specifies an asymptotic regime in which both the number of spots $n$ and the number of genes $d$ tend to infinity, with the remaining dimensions growing at most exponentially in the other.

\subsection{Theoretical Properties of VEM Estimators}

The traditional misclassification rate is defined as the minimum proportion of mislabeled observations over all permutations of the class labels:
\begin{equation*}
    \mathrm{MCR} = \min_{p(\cdot)} \frac{1}{n}\sum_{i=1}^n \mathbbm{1}\bigl(z_i^* \neq p(\widehat{z}_i)\bigr),
\end{equation*}
where $p(\cdot)$ is a permutation map. 
However, this definition is not appropriate when the number of estimated clusters exceeds the true number.
Inspired by \cite{zhang2021label}, we introduce an extended notion of misclassification rate that applies to soft assignments (probability matrices) and handles the overspecified case.

\begin{definition}
  Let $\mathcal{G}_r$ be the set of all surjective maps $g_r: [K] \to [K_0]$, and let $\mathbf{I}^{\bm z^*}$ be the true row-cluster indicator matrix with $I^{\bm z^*}_{ik}=z^*_{ik}$.
  The generalized row misclassification rate between $\mathbf{I}^{\bm z^*}$ and a probability matrix $\bm s$ is defined as
  \begin{equation*}
      \mathrm{M}_{\mathrm{row}}(\bm s) = \min_{g_r\in\mathcal{G}_r}\Bigl[ 1 - \sum_{k=1}^{K_0}\sum_{k'\in g_r^{-1}(k)} R_{kk'}(\mathbf{I}^{\bm z^*},\bm s) \Bigr],
  \end{equation*}
  where $g_r^{-1}(k)=\{k'\in[K]: g_r(k')=k\}$, and $R_{kk'}(\mathbf{I}^{\bm z^*},\bm s) = \frac{1}{n}\sum_{i=1}^{n} I^{\bm z^*}_{ik} s_{ik'}$.
  The generalized column misclassification rate $\mathrm{M}_{\mathrm{col}}(\bm t)$ is defined analogously using surjection $g_c: [L] \to [L_0]$.
\end{definition}

\noindent\textbf{Toy example.}
Let the true label vector be $\bm z^*=(1,1,2,2)$, and suppose we obtain soft assignments $\bm s$ under overspecification with $K=4$. Consider the surjection $g_r$ such that $g_r(1)=g_r(2)=1$ and $g_r(3)=g_r(4)=2$. Then the generalized row misclassification rate evaluates to $\mathrm{M}_{\mathrm{row}}(\bm s)=0.1$:
\[
\mathbf{I}^{\bm z^*} = 
\begin{bmatrix}
1 & 0\\
1 & 0\\
0 & 1\\
0 & 1
\end{bmatrix},\quad
\bm s = 
\begin{bmatrix}
0.3 & 0.6 & 0.01 & 0.09\\
0.5 & 0.4 & 0.09 & 0.01\\
0.01& 0.09& 0.3  & 0.6\\
0.09& 0.01& 0.5  & 0.4
\end{bmatrix}
\xrightarrow{g_r}
\begin{bmatrix}
0.9 & 0.1\\
0.9 & 0.1\\
0.1 & 0.9\\
0.1 & 0.9
\end{bmatrix}.
\]

Let $(\widehat{\bm s}, \widehat{\bm t}, \widehat{\bm\Theta}) = \arg\max \mathcal{J}(\bm s, \bm t, \bm\Theta)$ and define the overall generalized misclassification rate $\widehat{\mathrm{M}} = \max\bigl(\mathrm{M}_{\mathrm{row}}(\widehat{\bm s}), \mathrm{M}_{\mathrm{col}}(\widehat{\bm t})\bigr)$.

\begin{theorem}\label{VEM_theory}
  Under Conditions $(\mathrm{C}1)$, $(\mathrm{C}2)$ and $(\mathrm{C}3)$, the following results hold:
  \begin{enumerate}[(i)]
      \item For any $\delta>0$, $\widehat{\mathrm{M}} = o_p\bigl((nd)^{-\delta}\bigr)$.
      
      \item There exists a constant $0<c<1/2$ such that
      $\mathbb{P}\bigl( cn \le \sum_{i=1}^n \widehat{s}_{ik} \le (1-c)n \bigr) \to 1$ for all $k\in[K]$, and
      $\mathbb{P}\bigl( cd \le \sum_{j=1}^d \widehat{t}_{jl} \le (1-c)d \bigr) \to 1$ for all $l\in[L]$.

      \item Let $g_r$ and $g_c$ be the surjective mappings that minimize the misclassification rate. Then for any $k'\in g_r^{-1}(k)$ and $l'\in g_c^{-1}(l)$,
      $\sqrt{nd}\,\bigl(\widehat{\mu}_{k'l'} - \mu^*_{kl}\bigr) = O_p(1)$.
  \end{enumerate}
\end{theorem}

Theorem~\ref{VEM_theory}(i) shows that the generalized misclassification rate decays faster than any polynomial rate in the total sample size $nd$, implying that spurious cluster splitting disappears asymptotically.
Theorem~\ref{VEM_theory}(ii) further guarantees that, despite the absence of observed labels, no estimated cluster degenerates to a vanishing proportion; every working cluster retains a non-negligible mass. 
Finally, Theorem~\ref{VEM_theory}(iii) demonstrates that the VEM estimates of the block-specific means are $\sqrt{nd}$-consistent: up to the fusion of redundant clusters, the estimated means converge to the true values at the optimal rate.

\subsection{Theoretical Properties of Penalized Fusion}
To evaluate clustering accuracy after fusion, we define the overall generalized misclassification rate based on the merged variational posteriors: $\widetilde{\mathrm{M}} = \max\bigl(\mathrm{M}_{\mathrm{row}}(\widetilde{\bm s}), \mathrm{M}_{\mathrm{col}}(\widetilde{\bm t})\bigr)$. We next demonstrate the theoretical properties of the penalized fusion procedure.
\begin{theorem}\label{The_properties_of_penalized_fusion_algorithm}

  Let the adaptive weights be constructed as in \eqref{weights} using the initial VEM estimator $\widehat{\bm\mu}$.
  Assume the tuning parameter $nd\gamma_{nd}\to\infty$ and $\sqrt{nd}\gamma_{nd}\to0$ as $n,d\to\infty$.
  With the same surjective maps $g_r$ and $g_c$ as in Theorem~\ref{VEM_theory}, and under Conditions $(\mathrm{C}1)$, $(\mathrm{C}2)$ and $(\mathrm{C}3)$, we have:
    \begin{enumerate}[(i)]
    \item $\mathbb{P}(\widetilde{K} = K_0, \widetilde{L}=L_0)\to1$ and for any $\delta>0$, $\widetilde{\mathrm{M}} = o_p\bigl((nd)^{-\delta}\bigr)$.
    \item For $k' = g_r^{-1}(k)$ and $l'=g_c^{-1}(l)$, $\sqrt{nd}(\tilde{\mu}_{k'l'} - \mu_{kl}^*) \xrightarrow{d} N(0, (\pi_k^*\rho_l^*)^{-1}\lambda_{kl}^*)$.
    \end{enumerate}
\end{theorem}
Theorem~\ref{The_properties_of_penalized_fusion_algorithm} establishes that the penalized fusion procedure consistently recovers the true co-cluster partition and yields asymptotically normal estimators for the block-specific means.
Together with Theorem~\ref{VEM_theory}, a single iteration of the fusion step on $\bm h$ and $\bm v$ is asymptotically equivalent to the fully iterated solution. 
This equivalence accounts for the substantial computational savings achieved by the penalized fusion method.

\subsection{Improvement by Introducing Spatial Smoothness}

In this subsection, we quantify the contribution of the Potts model to clustering accuracy with the assumption that $K_0$ and $L_0$ are known.  
Let $\bm x_i$ denote the observation at spot $i$. 
The classification rule assigns $\widehat{z}_i = \arg\max_k \pi_k A_{ik}$. 
Because the tails of the Gaussian density decay exponentially, in high-dimensional settings the misclassification probability between cluster $k_1$ and its nearest cluster $k_2$ dominates the misclassification probabilities against all other clusters. 
In the non-spatial case, the probability that an observation from cluster $k_1$ is incorrectly assigned to cluster $k_2$ is 
\begin{equation}\label{nonspatial_prob}
    \mathbb{P}\left( \widehat{z}_i = k_2 \mid z_i^* = k_1 \right)
    = \mathbb{P}\left( \frac{A_{i,k_1}}{A_{i,k_2}} < \frac{\pi_{k_2}}{\pi_{k_1}} \mid z_i^* = k_1 \right),
\end{equation}
and in the spatial case, the corresponding probability becomes
\begin{equation}\label{spatial_prob}
    \mathbb{P}_s\bigl( \widehat{z}_i = k_2 \mid z_i^* = k_1 \bigr)
    = \mathbb{P}\left( \frac{A_{i,k_1}}{A_{i,k_2}} < \frac{\pi_{i,k_2}}{\pi_{i,k_1}} \mid z_i^* = k_1 \right),
\end{equation}
where $\mathbb{P}$ and $\mathbb{P}_s$ denote probabilities under the non-spatial and spatial models, respectively.
Comparing \eqref{nonspatial_prob} and \eqref{spatial_prob},
we see that the non-spatial misclassification rate depends on the ratio of global mixing proportions $\pi_{k_2}/\pi_{k_1}$, whereas the spatial misclassification rate depends on the ratio of local prior probabilities
\[
    \frac{\pi_{i,k_2}}{\pi_{i,k_1}}
    = \exp\Bigl\{ \beta\sum_{j\in\mathcal{N}_i}\bigl[\mathbbm{1}(z_j=k_2)-\mathbbm{1}(z_j = k_1)\bigr] \Bigr\}.
\]
When $\beta>0$, neighboring spots tend to share the same cluster label, which shrinks the ratio $\pi_{i,k_2}/\pi_{i,k_1}$ and thereby facilitates correct classification.

Few works rigorously quantify how spatial correlation reduces misclassification rates in spatial transcriptomics clustering. 
We quantify the clustering gains from incorporating the Potts model under a disordered phase and a homoscedastic data assumption. 
The disordered phase ensures ergodicity of the Potts model and excludes symmetry breaking, thereby simplifying the asymptotic analysis \citep{baxter1985exactly}.
The homoscedasticity assumption is motivated by two considerations. 
First, normalization and log-transformation of UMI counts form the standard preprocessing pipeline for ST data. 
This pipeline is explicitly designed to stabilize variance, which reduces the strong dependence of gene-expression variability on the mean. 
Second, the homoscedasticity assumption makes the analysis of Gaussian likelihood ratios much simpler.

\begin{proposition}\label{Spatial_smoothness}
      Suppose the system of Potts model is in the disordered phase and the covariance structure is homogeneous across row clusters. Under Conditions $(\mathrm{C}1)$, $(\mathrm{C}2)$ and $(\mathrm{C}3)$, we have
  \begin{equation*}
    \frac{\mathbb{P}_s\left\{ \widehat{z}_i \neq z_i^* \right\}}{\mathbb{P}\left\{ \widehat{z}_i \neq z_i^* \right\}} \le \left( \frac{2e^{\beta/2}+K-2}{e^{\beta}+K-1}\right)^{|\mathcal{N}_i|} + o(1).
\end{equation*}
\end{proposition}
Proposition~\ref{Spatial_smoothness} states that incorporating a Potts-type spatial prior reduces the spot-level misclassification probability relative to the non-spatial rule, and that this reduction strengthens monotonically as the spatial smoothing parameter $\beta$ increases.
By shrinking the local prior odds toward the dominant label within each neighborhood, the Potts model creates locally favorable conditions for correct classification and, consequently, lowers the overall misclassification rate of the clustering procedure.

\section{Simulation}\label{Simulation}
\subsection{Data Generation}

In all experiments, we generate datasets of size $(n,d) = (529, 500)$ on a $23 \times 23$ rectangular lattice with unit spacing ($r=1$). 
The true numbers of row and column clusters are fixed at $K_0 = L_0 = 3$, and the column-cluster mixing proportions are set to $\bm\rho = (0.25, 0.35, 0.4)^\top$.
Let $\bm\xi_i = (\xi_{i1}, \xi_{i2})^\top$ denote the spatial coordinates of spot $i$, and let $\mathrm{len} = 23$ be the length of the lattice.

\begin{figure}[htbp]
    \centering
    \includegraphics[width=0.6\textwidth]{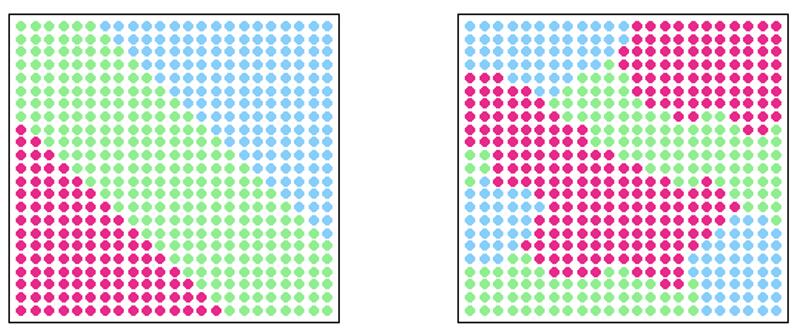} 
    \caption{Spatial topologies for Case 1 (left) and Case 2 (right).}
    \label{fig:generate_z}
\end{figure}

We consider two mechanisms for generating the row-cluster labels $\bm z$.
The first mechanism, denoted by $p_1(\bm z)$, partitions the lattice using the linear boundaries
\[
  z_i = 3 - 2 \times \mathbbm{1}\bigl(\xi_{i1} + \xi_{i2} < 0.7 \cdot \mathrm{len}\bigr) - \mathbbm{1}\bigl(\xi_{i1} + \xi_{i2} > 1.3 \cdot \mathrm{len}\bigr),
\]
yielding the deterministic spatial topology shown in the left panel of Figure~\ref{fig:generate_z}.
The second mechanism, denoted by $p_2(\bm z)$, draws $\bm z$ from the Potts model \eqref{Potts} via Gibbs sampling, producing the stochastic topology shown in the right panel of Figure~\ref{fig:generate_z}.

We design four simulation cases that collectively reflect typical challenges in spatial transcriptomics analysis. 
Case 1 serves as the baseline. Data are generated from the Gaussian LBM with $\bm z \sim p_1(\bm z)$, $\bm\mu = (-1,1,0;0,-1,1;1,0,-1)$ and $\bm\Lambda = (40,50,60;40,50,60;40,50,60)$. 
Case 2 evaluates robustness under more irregular spatial boundaries. The model parameters remain the same as in Case 1, but the row labels are generated from the Potts model, $\bm z \sim p_2(\bm z)$. This case evaluates robustness against more irregular spatial boundaries. 
Case 3 examines low signal-to-noise ratio conditions. We weaken cluster separation by shrinking the block means. The row labels follow $p_1(\bm z)$, $\bm\Lambda$ is unchanged, and $\bm\mu = (-0.7,0.7,0;0,-0.7,0.7;0.7,0,-0.7)$. 
Case 4 assesses robustness to non-Gaussianity. Row labels are drawn from $p_1(\bm z)$, and raw counts are first sampled from a Negative Binomial distribution with mean matrix $\bm\mu_{\text{NB}} = (10,15,20;15,20,10;20,10,15)$ and common dispersion parameter $\phi_{kl} = 1$. Following standard spatial transcriptomics preprocessing, the simulated counts are then normalized and log-transformed before model fitting.

\subsection{Results}\label{Results}

We compare the proposed method, Fused Spatial LBM (F-SpLBM), against several alternatives, including F-LBM, LBM, SpLBM, the SpaRTaCo model \citep{sottosanti2023co}, a naive independent K-means approach (applying K-means separately to spots and genes), and spectral co-clustering \citep{battaglia2024co}. 
For LBM and SpLBM, $K$ and $L$ are selected via the ICL. 
For the last three methods, $K$ and $L$ are provided to ensure a fair comparison of clustering performance.

We evaluate clustering accuracy using the adjusted Rand index (ARI) and normalized mutual information (NMI). Larger values show better agreement with the true partition. 
We also compute Moran's I and Geary's C indices to quantify spatial coherence. 
Higher Moran’s I and lower Geary’s C indicate greater spatial smoothness. 
More details on these metrics can be found in \cite{yuan2024benchmarking}. 
All reported results are averaged over 50 independent simulation replicates.
The results are summarized in Tables \ref{tab:simulation1}-\ref{tab:simulation4}.
Overall, F-SpLBM attains the best clustering accuracy and spatial smoothness across all four scenarios. 
A detailed inspection of the results reveals several clear patterns.

The SpLBM and standard LBM rely on the ICL for model selection and tend to overestimate $K$ and $L$. 
In contrast, the fused variants (F-SpLBM and F-LBM) produce estimates that are tightly concentrated around the true values. 
This confirms that the penalized fusion strategy effectively identifies the correct co-cluster structure by merging redundant centers. 
This advantage is maintained under low signal-to-noise conditions (Case 3) and model misspecification (Case 4).

The fusion-based methods achieve higher ARI and NMI with smaller or comparable standard deviations than their unfused counterparts (SpLBM and LBM). 
This shows gains in both accuracy and estimation stability. 
Moreover, the spatial methods consistently outperform their non-spatial counterparts: the F-SpLBM surpasses the F-LBM, and the SpLBM surpasses the LBM, especially on the row-clustering metrics.
This confirms that the Potts-model provides additional discriminative power beyond the transcriptional information alone. 
Although SpaRTaCo also incorporates spatial dependence through a covariance structure, it performs worse than several competitors. 
This suggests that modeling spatial correlations through a latent covariance matrix may needlessly complicate the model and lower clustering accuracy.

A comparison of Case 1 and Case 2 shows that the performance of both the F-SpLBM and the SpLBM remains stable when the spatial boundaries become more irregular. 
This indicates that the Potts model is robust to the complexity of the spatial topology.
The row-level ARI of the F-SpLBM, F-LBM, SpLBM, and LBM decreases by 0.11, 0.24, 0.20, and 0.27, respectively, from Case 1 to Case 3. 
In Case 3, the fusion-based methods show a much smaller drop in performance and barely any increase in standard deviation, whereas the unfused methods experience a sharp increase in variability. 
This contrast shows both the stronger resilience to weak cluster separation and the better stability of penalized fusion in low-signal settings.
In Case 4, the F-SpLBM continues to perform competitively, achieving the highest row ARI (0.98) and strong spatial smoothness. 
Nevertheless, the F-SpLBM still delivers the most stable and consistently superior spatial smoothness, supporting the adequacy of the Gaussian working model after standard preprocessing of ST data.

{
% ========== 核心优化：字体大小 + 列间距 + 行间距 ==========
\small          % 字体缩小至7pt（比\footnotesize更小，核心压缩）
\setlength{\tabcolsep}{2.5pt}  % 列间距缩至1.5pt（默认6pt，极致压缩）

\begin{longtable}{@{}ccccccccc@{} % 消除表格左侧默认空白
  >{\RaggedRight\arraybackslash}p{1cm}  % Model列：固定宽+左对齐（避免溢出）
  >{\Centering\arraybackslash}p{1cm}    % K列：固定宽+居中
  >{\Centering\arraybackslash}p{1cm}    % L列：固定宽+居中
  >{\Centering\arraybackslash}p{1cm}    % row ARI
  >{\Centering\arraybackslash}p{1cm}    % row NMI
  >{\Centering\arraybackslash}p{1cm}    % col ARI
  >{\Centering\arraybackslash}p{1cm}    % col NMI
  >{\Centering\arraybackslash}p{1cm}    % Moran
  >{\Centering\arraybackslash}p{1cm}    % Geary
  @{} % 消除表格右侧默认空白
} % 5列自适应宽度，无默认边距

\caption{Clustering accuracy and spatial smoothness for Case 1. Mean (SD)}
\label{tab:simulation1}\tabularnewline
\toprule
Model &K &L &row ARI &row NMI &col ARI &col NMI &Moran &Geary  \\
\midrule
\endfirsthead

% 跨页后的表头
\toprule
Model &K &L &row ARI &row NMI &col ARI &col NMI &Moran &Geary  \\
\midrule
\endhead

% 最后一页的页脚
\bottomrule
\endlastfoot

% 第一组数据
{F-SpLBM}  &\textbf{3.08}$_{(0.27)}$ &\textbf{3.06}$_{(0.24)}$ &\textbf{0.97}$_{(0.05)}$ &\textbf{0.97}$_{(0.03)}$ &\textbf{0.97}$_{(0.04)}$ &\textbf{0.96}$_{(0.02)}$ &\textbf{0.89}$_{(0.02)}$ &\textbf{0.11}$_{(0.03)}$ \\
F-LBM &\textbf{3.08}$_{(0.27)}$ &\textbf{3.06}$_{(0.24)}$ &$0.92_{(0.05)}$ &$0.88_{(0.03)}$ &\textbf{0.97}$_{(0.04)}$ &\textbf{0.96}$_{(0.02)}$  &$0.82_{(0.03)}$ &$0.17_{(0.03)}$ \\
SpLBM  &$3.48_{(0.68)}$ &$3.94_{(0.77)}$ &$0.96_{(0.06)}$ &$0.95_{(0.04)}$ &$0.92_{(0.07)}$ &$0.92_{(0.05)}$  &$0.85_{(0.08)}$ &$0.14_{(0.09)}$ \\
LBM &$4.00_{(0.81)}$ &$3.86_{(0.73)}$ &$0.87_{(0.08)}$ &$0.84_{(0.05)}$ &$0.92_{(0.07)}$ &$0.93_{(0.05)}$  &$0.70_{(0.17)}$ &$0.29_{(0.17)}$ \\
SpaRTaCo  &- &- &$0.41_{(0.15)}$ &$0.39_{(0.12)}$ &$0.89_{(0.06)}$ &$0.86_{(0.06)}$ &$0.34_{(0.18)}$ &$0.66_{(0.18)}$\\
K-means &- &- &$0.80_{(0.10)}$ &$0.74_{(0.08)}$ &$0.75_{(0.12)}$ &$0.72_{(0.09)}$  &$0.71_{(0.10)}$ &$0.29_{(0.10)}$ \\
Spectral &- &- &$0.71_{(0.19)}$ &$0.65_{(0.15)}$ &$0.74_{(0.13)}$ &$0.70_{(0.10)}$ &$0.64_{(0.15)}$ &$0.36_{(0.15)}$ \\

\end{longtable}
}

{
% ========== 核心优化：字体大小 + 列间距 + 行间距 ==========
\small          % 字体缩小至7pt（比\footnotesize更小，核心压缩）
\setlength{\tabcolsep}{2.5pt}  % 列间距缩至1.5pt（默认6pt，极致压缩）

\begin{longtable}{@{}ccccccccc@{} % 消除表格左侧默认空白
  >{\RaggedRight\arraybackslash}p{1cm}  % Model列：固定宽+左对齐（避免溢出）
  >{\Centering\arraybackslash}p{1cm}    % K列：固定宽+居中
  >{\Centering\arraybackslash}p{1cm}    % L列：固定宽+居中
  >{\Centering\arraybackslash}p{1cm}    % row ARI
  >{\Centering\arraybackslash}p{1cm}    % row NMI
  >{\Centering\arraybackslash}p{1cm}    % col ARI
  >{\Centering\arraybackslash}p{1cm}    % col NMI
  >{\Centering\arraybackslash}p{1cm}    % Moran
  >{\Centering\arraybackslash}p{1cm}    % Geary
  @{} % 消除表格右侧默认空白
} % 5列自适应宽度，无默认边距

\caption{Clustering accuracy and spatial smoothness for Case 2. Mean (SD)}
\label{tab:simulation2}\tabularnewline
\toprule
Model &K &L &row ARI &row NMI &col ARI &col NMI &Moran &Geary  \\
\midrule
\endfirsthead

% 跨页后的表头
\toprule
Model &K &L &row ARI &row NMI &col ARI &col NMI &Moran &Geary  \\
\midrule
\endhead

% 最后一页的页脚
\bottomrule
\endlastfoot

% 第二组数据
{F-SpLBM}  &\textbf{3.08}$_{(0.27)}$ &\textbf{3.02}$_{(0.14)}$ &\textbf{0.97}$_{(0.06)}$ &\textbf{0.96}$_{(0.04)}$ &\textbf{0.99}$_{(0.02)}$ &\textbf{0.98}$_{(0.02)}$  &\textbf{0.78}$_{(0.06)}$ &\textbf{0.21}$_{(0.06)}$  \\
F-LBM &\textbf{3.08}$_{(0.27)}$ &\textbf{3.02}$_{(0.14)}$ &$0.92_{(0.06)}$ &$0.88_{(0.04)}$ &\textbf{0.99}$_{(0.02)}$ &\textbf{0.98}$_{(0.02)}$ &$0.73_{(0.09)}$ &$0.27_{(0.09)}$ \\
SpLBM  &$3.50_{(0.71)}$ &$3.96_{(0.81)}$ &$0.95_{(0.08)}$ &$0.94_{(0.05)}$ &$0.92_{(0.08)}$ &$0.94_{(0.05)}$  &$0.74_{(0.10)}$ &$0.25_{(0.11)}$\\
LBM &$3.94_{(0.77)}$ &$3.74_{(0.69)}$ &$0.86_{(0.09)}$ &$0.84_{(0.05)}$ &$0.93_{(0.08)}$ &$0.94_{(0.05)}$  &$0.62_{(0.17)}$ &$0.37_{(0.17)}$ \\
SpaRTaCo &- &- &$0.53_{(0.13)}$ &$0.50_{(0.11)}$ &$0.96_{(0.03)}$ &$0.94_{(0.04)}$  &$0.41_{(0.14)}$ &$0.59_{(0.14)}$  \\
K-means &- &- &$0.78_{(0.16)}$ &$0.72_{(0.12)}$ &$0.71_{(0.17)}$ &$0.70_{(0.12)}$  &$0.61_{(0.13)}$ &$0.38_{(0.13)}$ \\
Spectral &- &- &$0.72_{(0.17)}$ &$0.66_{(0.14)}$ &$0.71_{(0.14)}$ &$0.68_{(0.11)}$  &$0.57_{(0.16)}$ &$0.43_{(0.16)}$\\

\end{longtable}
}

{
% ========== 核心优化：字体大小 + 列间距 + 行间距 ==========
\small          % 字体缩小至7pt（比\footnotesize更小，核心压缩）
\setlength{\tabcolsep}{2.5pt}  % 列间距缩至1.5pt（默认6pt，极致压缩）

\begin{longtable}{@{}ccccccccc@{} % 消除表格左侧默认空白
  >{\RaggedRight\arraybackslash}p{1cm}  % Model列：固定宽+左对齐（避免溢出）
  >{\Centering\arraybackslash}p{1cm}    % K列：固定宽+居中
  >{\Centering\arraybackslash}p{1cm}    % L列：固定宽+居中
  >{\Centering\arraybackslash}p{1cm}    % row ARI
  >{\Centering\arraybackslash}p{1cm}    % row NMI
  >{\Centering\arraybackslash}p{1cm}    % col ARI
  >{\Centering\arraybackslash}p{1cm}    % col NMI
  >{\Centering\arraybackslash}p{1cm}    % Moran
  >{\Centering\arraybackslash}p{1cm}    % Geary
  @{} % 消除表格右侧默认空白
} % 5列自适应宽度，无默认边距

\caption{Clustering accuracy and spatial smoothness for Case 3. Mean (SD)}
\label{tab:simulation3}\tabularnewline
\toprule
Model &K &L &row ARI &row NMI &col ARI &col NMI &Moran &Geary  \\
\midrule
\endfirsthead

% 跨页后的表头
\toprule
Model &K &L &row ARI &row NMI &col ARI &col NMI &Moran &Geary  \\
\midrule
\endhead

% 最后一页的页脚
\bottomrule
\endlastfoot

% 第三组数据
{F-SpLBM}  &\textbf{3.10}$_{(0.30)}$ &\textbf{3.10}$_{(0.30)}$ &\textbf{0.86}$_{(0.07)}$ &\textbf{0.81}$_{(0.06)}$ &\textbf{0.91}$_{(0.05)}$ &\textbf{0.88}$_{(0.04)}$  &\textbf{0.80}$_{(0.04)}$ &\textbf{0.19}$_{(0.04)}$ \\
F-LBM &\textbf{3.10}$_{(0.30)}$ &\textbf{3.10}$_{(0.30)}$ &$0.68_{(0.07)}$ &$0.60_{(0.05)}$ &\textbf{0.91}$_{(0.05)}$ &\textbf{0.88}$_{(0.04)}$  &$0.60_{(0.05)}$ &$0.39_{(0.05)}$ \\
SpLBM &$4.06_{(0.77)}$ &$3.96_{(0.73)}$ &$0.76_{(0.13)}$ &$0.73_{(0.09)}$ &$0.87_{(0.07)}$ &$0.85_{(0.05)}$  &$0.65_{(0.16)}$ &$0.35_{(0.16)}$  \\
LBM &$4.32_{(0.74)}$ &$3.86_{(0.76)}$ &$0.60_{(0.10)}$ &$0.55_{(0.06)}$ &$0.87_{(0.06)}$ &$0.86_{(0.05)}$  &$0.46_{(0.16)}$ &$0.54_{(0.17)}$ \\
SpaRTaCo &- &- &$0.16_{(0.09)}$ &$0.15_{(0.08)}$ &$0.74_{(0.11)}$ &$0.72_{(0.09)}$  &$0.12_{(0.11)}$ &$0.88_{(0.11)}$  \\
K-means &- &- &$0.11_{(0.07)}$ &$0.11_{(0.06)}$ &$0.20_{(0.08)}$ &$0.22_{(0.07)}$  &$0.10_{(0.08)}$ &$0.90_{(0.09)}$ \\
Spectral &- &- &$0.16_{(0.07)}$ &$0.16_{(0.06)}$ &$0.23_{(0.07)}$ &$0.24_{(0.07)}$  &$0.14_{(0.09)}$ &$0.86_{(0.09)}$ \\

\end{longtable}
}

{
% ========== 核心优化：字体大小 + 列间距 + 行间距 ==========
\small          % 字体缩小至7pt（比\footnotesize更小，核心压缩）
\setlength{\tabcolsep}{2.5pt}  % 列间距缩至1.5pt（默认6pt，极致压缩）

\begin{longtable}{@{}ccccccccc@{} % 消除表格左侧默认空白
  >{\RaggedRight\arraybackslash}p{1cm}  % Model列：固定宽+左对齐（避免溢出）
  >{\Centering\arraybackslash}p{1cm}    % K列：固定宽+居中
  >{\Centering\arraybackslash}p{1cm}    % L列：固定宽+居中
  >{\Centering\arraybackslash}p{1cm}    % row ARI
  >{\Centering\arraybackslash}p{1cm}    % row NMI
  >{\Centering\arraybackslash}p{1cm}    % col ARI
  >{\Centering\arraybackslash}p{1cm}    % col NMI
  >{\Centering\arraybackslash}p{1cm}    % Moran
  >{\Centering\arraybackslash}p{1cm}    % Geary
  @{} % 消除表格右侧默认空白
} % 5列自适应宽度，无默认边距

\caption{Clustering accuracy and spatial smoothness for Case 4. Mean (SD)}
\label{tab:simulation4}\tabularnewline
\toprule
Model &K &L &row ARI &row NMI &col ARI &col NMI &Moran &Geary  \\
\midrule
\endfirsthead

% 跨页后的表头
\toprule
Model &K &L &row ARI &row NMI &col ARI &col NMI &Moran &Geary  \\
\midrule
\endhead

% 最后一页的页脚
\bottomrule
\endlastfoot

{F-SpLBM}  &\textbf{3.16}$_{(0.37)}$ &\textbf{3.22}$_{(0.46)}$ &\textbf{0.98}$_{(0.06)}$ &\textbf{0.98}$_{(0.04)}$ &\textbf{0.96}$_{(0.08)}$ &\textbf{0.98}$_{(0.04)}$  &\textbf{0.85}$_{(0.12)}$ &\textbf{0.14}$_{(0.13)}$  \\
F-LBM &\textbf{3.16}$_{(0.37)}$ &\textbf{3.22}$_{(0.46)}$ &$0.97_{(0.07)}$ &\textbf{0.98}$_{(0.04)}$ &\textbf{0.96}$_{(0.08)}$ &\textbf{0.98}$_{(0.04)}$  &$0.84_{(0.18)}$ &$0.16_{(0.18)}$ \\
SpLBM &$3.84_{(0.91)}$ &$4.66_{(0.52)}$ &$0.92_{(0.11)}$ &$0.95_{(0.06)}$ &$0.81_{(0.09)}$ &$0.88_{(0.04)}$  &$0.82_{(0.13)}$ &$0.18_{(0.13)}$  \\
LBM &$4.38_{(0.75)}$ &$4.40_{(0.73)}$ &$0.84_{(0.11)}$ &$0.90_{(0.06)}$ &$0.84_{(0.10)}$ &$0.90_{(0.06)}$  &$0.68_{(0.22)}$ &$0.31_{(0.22)}$  \\
SpaRTaCo &- &- &$0.64_{(0.20)}$ &$0.66_{(0.15)}$ &\textbf{0.96}$_{(0.06)}$ &$0.95_{(0.07)}$  &$0.57_{(0.22)}$ &$0.43_{(0.22)}$  \\
K-means &- &- &$0.96_{(0.14)}$ &$0.97_{(0.09)}$ &$0.91_{(0.20)}$ &$0.95_{(0.13)}$  &$0.87_{(0.13)}$ &$0.12_{(0.13)}$ \\
Spectral &- &- &$0.92_{(0.19)}$ &$0.95_{(0.12)}$ &$0.93_{(0.19)}$ &$0.96_{(0.12)}$  &$0.83_{(0.22)}$ &$0.16_{(0.22)}$ \\

\end{longtable}
}

\section{Real Data Analysis}\label{Real_Data}

We analyze a human dorsolateral prefrontal cortex (DLPFC) sample (subject 151673) 
from the spatial transcriptomics dataset generated by the Lieber Institute for 
Brain Development (LIBD) and originally studied by \cite{maynard2021transcriptome}. 
The data are publicly available through the \texttt{R} package \texttt{spatialLIBD} 
\citep{pardo2022spatiallibd}. 
It contains expression measurements for 33,538 genes across 1,639 spatially resolved spots, with gene expression quantified using UMI counts.
Following standard preprocessing procedures, we first apply the gene-selection procedure proposed by \cite{townes2019feature} to identify informative features. 
Specifically, we retain the top 500 genes with the highest variability and discard the remaining genes with limited signal. 
The raw UMI counts are then normalized and log-transformed to yield the final dataset used in the downstream analysis.

The penalized fusion mechanism in both the F-SpLBM and the F-LBM identifies $(K,L)=(6,4)$. 
The SpaRTaCo selects $(K,L)=(9,2)$ by maximizing the ICL \citep{sottosanti2023co}.
For the K-means method and the spectral co-clustering, we also adopt $(K, L) = (7, 4)$. Quantitative performance comparisons are reported in Table \ref{tab:real_data}.

\begin{longtable}[]{@{}ccccccc@{}}
\caption{Clustering performance of accuracy and spatial smoothness.}
\label{tab:real_data}\tabularnewline
\toprule\noalign{}
 Model &ARI &NMI &Moran &Geary \\
\midrule\noalign{}
\endfirsthead
\toprule\noalign{}
 Model &ARI &NMI &Moran &Geary \\
\midrule\noalign{}
\endhead
\bottomrule\noalign{}
\endlastfoot
F-SpLBM &0.53 &0.54 &0.65 &0.36\\
F-LBM &0.43 &0.46 &0.46 &0.55\\
Spartaco  &0.11  &0.23 &0.40 &0.61 \\
K-means &0.23 &0.39 &0.42 &0.58\\
Spectral & 0.22 & 0.33 &0.20 &0.80 \\
\end{longtable}

\begin{figure}[htbp]
    \centering
    \includegraphics[width=0.8\textwidth]{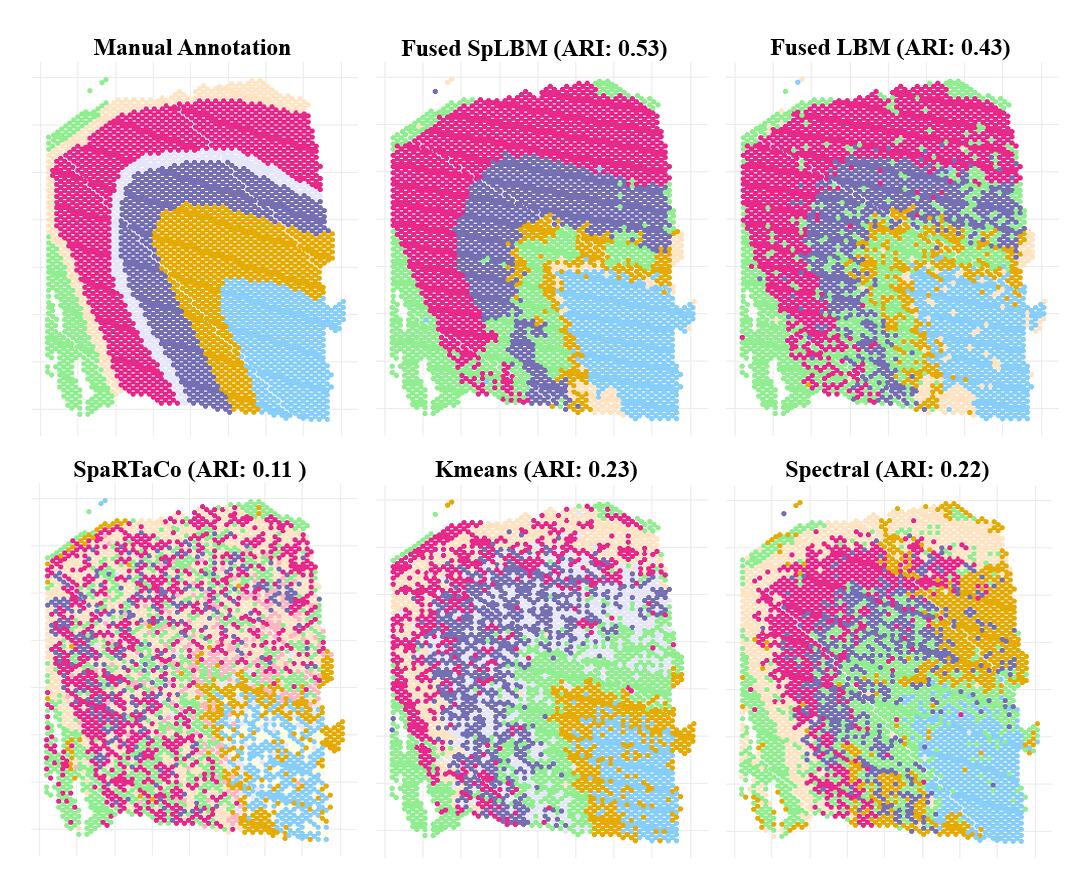} 
    \caption{Results on the human dorsolateral prefrontal cortex data.}
    \label{fig:real_data_res_plot}
\end{figure}

Figure~\ref{fig:real_data_res_plot} displays the clustering results of all competing methods. Compared with the alternatives, our method yields noticeably smoother spatial partitions that align more closely with the known laminar organization of the cortex. 

\begin{figure}[htbp]
    \centering
    \includegraphics[width=0.8\textwidth]{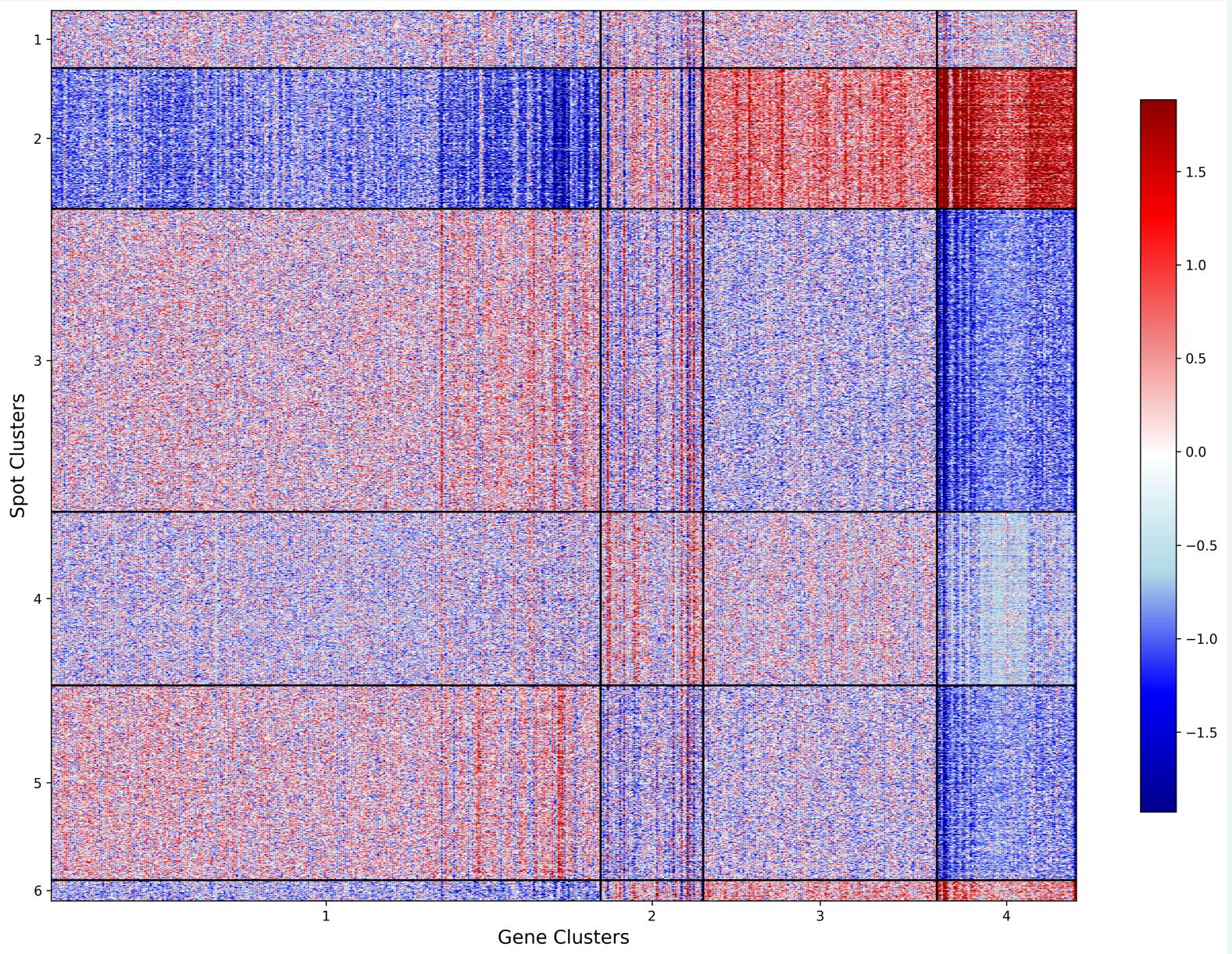} 
    \caption{The heatmap of data across different blocks detected by F-SpLBM.}
    \label{fig:heatmap}
\end{figure}

Figure~\ref{fig:heatmap} shows a heatmap of the $6\times4$ bolck structure identified by F-SpLBM. 
Several gene modules display marked overexpression or underexpression within specific spatial domains, suggesting coordinated transcriptional programs linked to region-specific biology.

\begin{figure}[htbp]
    \centering
    \includegraphics[width=0.9\textwidth]{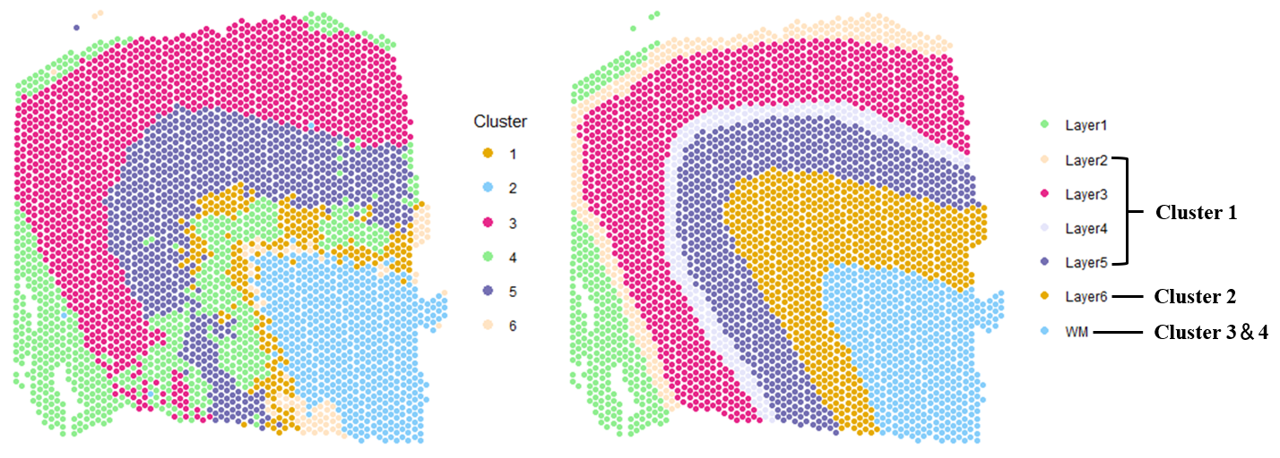} 
    \caption{Tissue locations where different gene clusters are active.}
    \label{fig:real_data_with_legend}
\end{figure}

The manual annotations delineate the white matter (WM) in the bottom-right portion of the tissue section, with cortical layers L6 through L1 arranged sequentially toward the top-left.
These annotations, together with the clustering results of the proposed method, are shown in Figure~\ref{fig:real_data_with_legend}. 
Upon comparison, we found that our clustering did not resolve independent clusters for layers L2 and L4, and there was substantial misclassification between L1 and L6. 
To explain this, we computed the block mean matrix based on the inferred gene cluster assignments and the true spot labels, and then calculated the Euclidean distances between the mean vectors of different true layers (Table~\ref{tab:layer_dist}). 
As shown, the distances for the pairs (L1, L6), and (L2, L3, L4, L5) are relatively small, making them difficult to separate, which is consistent with the observed clustering confusion.
\begin{longtable}[]{@{}r|rrrrrrr@{}}
\caption{The Euclidean distances between the mean vectors of different true layers.}
\label{tab:layer_dist}\tabularnewline
\toprule\noalign{}
 & Layer1 & Layer2 & Layer3 & Layer4 & Layer5 & Layer6 \\
\midrule\noalign{}
\endfirsthead
\toprule\noalign{}
 & Layer1 & Layer2 & Layer3 & Layer4 & Layer5 & Layer6 \\
\midrule\noalign{}
\endhead
\bottomrule\noalign{}
\endlastfoot
Layer2 & 0.49 &  &  &  &  & \\ 
Layer3 & 0.60 & 0.14 &  &  &  &  \\ 
Layer4 & 0.57 & 0.21 & 0.17 &  &  &  \\ 
Layer5 & 0.57 & 0.27 & 0.24 & 0.07 &  &  \\ 
Layer6 & 0.37 & 0.65 & 0.70 & 0.60 & 0.56 &  \\ 
WM    & 2.04 & 2.46 & 2.52 & 2.42 & 2.37 & 1.83 \\ 
\end{longtable}

Aligning the inferred spot clusters with these annotations reveals a clear correspondence between gene modules and cortical layers. 
Gene Cluster~1 (GC1) is predominantly expressed in Spot Cluster~3 (SC3) and SC5, which largely map to layers L2-L5. 
GC2 shows elevated expression mainly in SC4 and SC6, corresponding to L6. 
In contrast, GC3 and GC4 are primarily active in SC2 and SC6, both overlapping substantially with the annotated WM region. 
These findings demonstrate that our model captures complex spatial heterogeneity while yielding biologically interpretable co-clustering structures. 
The resulting spatially resolved gene modules offer meaningful guidance for downstream analyses such as functional enrichment and cell-type characterization.

We run Gene Ontology (GO) enrichment analysis on marker genes of each spatial domain using the \texttt{R} package \texttt{clusterProfiler} \citep{wu2021clusterprofiler}. 
Enrichment signals are strong across all clusters, with adjusted p-values from $10^{-4}$ to $10^{-14}$ for the representative terms.

GC1, characterized by representative genes such as {\it MAP1A} and {\it SLC17A7}, is enriched in synaptic functions, including modulation of chemical synaptic transmission, trans-synaptic signaling, and vesicle-mediated transport. 
This profile is concordant with the pyramidal-neuron-rich layers L2-L5, which mediate intracortical connectivity and information integration \citep{douglas2004neuronal}.
GC2, with {\it ND2} as a representative gene, is enriched in energy-derivation and oxidative phosphorylation pathways, consistent with the high metabolic demand of deep cortical layers processing thalamocortical input \citep{thomson2010neocortical}.
GC3, represented by ribosomal genes such as {\it RPLP1} and {\it RPS12}, is associated with cytoplasmic translation and ribosome assembly, indicating high-capacity protein synthesis.
GC4, represented by {\it GPM6B}, is linked to gliogenesis and axon ensheathment, pointing to structural and metabolic support of white matter integrity . 
Thus, GC3 and GC4 play complementary roles in white-matter maintenance \citep{fields2008white}.

In summary, the laminar organization of the DLPFC reflects pronounced functional specialization: middle layers support intensive synaptic processing and integration, deep layers sustain elevated metabolic activity, and white matter regions maintain structural integrity.
This spatial modularity of gene expression highlights the molecular basis of prefrontal cortex architecture and the biological interpretability of our co-clustering framework.

\section{Conclusion}

We have proposed F-SpLBM, a novel method that employs a penalized fusion strategy to efficiently determine the number of co-clusters and integrates a Potts model to capture the spatial smoothness.
Simulation studies and real-data applications demonstrate that our method effectively leverages spatial information, leading to improved clustering accuracy and enhanced spatial coherence of the identified tissue domains.
We establish that the fusion-based procedure recovers the true
block structure with the misclassification rate converging at a super-polynomial rate. 
We further prove the asymptotic normality of the parameter estimators and quantify the improvement in clustering accuracy brought by incorporating spatial smoothness. 

We connect the upper bound of the misclassification rate to the Kullback-Leibler (KL) divergence to establish its convergence. 
Besides, we decompose the VEM estimator into the MLE and an error term induced by clustering errors, thereby deriving asymptotic normality of the parameter estimators.
Both strategies can potentially be extended to other distribution assumptions, such as the Poisson distribution. 
In Proposition~\ref{Spatial_smoothness}, we adopt a homoscedasticity assumption to simplify the analysis of the likelihood ratio. 
This restriction may be relaxed in future work with the help of better mathematical tools. 
Finally, since the spatial smoothing step can be seen as a post-processing module, the proposed framework also applies to non-spatial datasets.

\phantomsection\label{supplementary_material}
\bigskip
\begin{center}
{\large\bf SUPPLEMENTARY MATERIAL}
\end{center}
The supplementary material contains the proofs of all the theoretical results. 
%The R code for simulation studies and real data analysis is also provided.

%\phantomsection\label{funding}
%\bigskip
%\begin{center}
%{\large\bf FUNDING}
%\end{center}
%This study is partially supported by the National Natural Science Foundation of China [12571313], the Shuimu Tsinghua Scholar Program of Tsinghua University, the China Postdoctoral Science Foundation (2025M773050), the Postdoctoral Fellowship Program (Grade B) of China Postdoctoral Science Foundation (GZB20250703).

\phantomsection\label{disclosure_statement}
\bigskip
\begin{center}
{\large\bf DISCLOSURE STATEMENT}
\end{center}
The authors report there are no competing interests to declare. The authors confirm that they have read and comply with the Taylor \& Francis AI policy. ChatGPT-4o was used to assist in language
refinement during manuscript preparation. All content was reviewed and verified by the authors.

\bibliographystyle{agsm}
\bibliography{main}
%\bibliography{ref}
\end{document}